\documentclass[twocolumn,aps,prl,showpacs,,showkeys,groupedaddress,amsmath,amssymb]{revtex4}
\usepackage{amsmath}
\usepackage{amssymb}
\usepackage{txfonts}
\usepackage{graphicx}
\usepackage{dcolumn}
\begin{document}
\title{Efficient Long-distance Quantum Communication Using Microtoroidal Resonators}
\author{Fang-Yu Hong}
\author{Shi-Jie Xiong }

\affiliation{National Laboratory of Solid State Microstructures and
Department of Physics, Nanjing University, Nanjing 210093, China}
\date{\today}
\begin{abstract}
Based on the
interaction between a three-level system and a microtoroidal
resonator, we present a scheme for long-distance quantum
communication in which entanglement generation with near 0.5 success
probability and swaps can be implemented by accurate state detection
via measuring about 100 photons. With this scheme the average time
of successful entanglement distribution over 2500 km with high
fidelity can be decreased to only about 30 ms, by 7 orders of
magnitude smaller compared with famous Duan-Lukin-Cirac-Zoller
(DLCZ) protocol [L.-M. Duan {\it et al.} Nature (London) {\bf414},
413 (2001)].
\end{abstract}



\maketitle

In quantum information science \cite{pzol}, the entanglement
distributed over quantum networks is a crucial requisite for
metrology \cite{vgsl}, quantum computation \cite{jcpz,ldhk}, and
communication \cite{jcpz,hbwd}. Quantum communication promises
completely secure transmission of keys with the Ekert protocol
\cite{aeke} and exact transfer of quantum states by quantum
teleportation \cite {chbe}. Because of losses and other noises in
quantum channels, the communication fidelity dwindles exponentially
with the channel length. In principle, this problem can be overcome
by applying quantum repeaters \cite{hbwd}, of which the basic
principle consists of separating the full distance into shorter
elementary links and entangling the links with quantum swaps
\cite{chbe,zzhe}. One kind of quantum repeater protocols based on single or sub-photon coherent state transmission may have high initial fidelity of entanglement but have low successful entangling event \cite{lcjt,dlcz}.  A protocol of specific importance for
long-distance quantum communication based on collective excitations
in atomic ensembles has been proposed in a seminal paper of Duan
{\it et al.} \cite {dlcz}. After that considerable efforts have been
devoted along this line \cite{crfp,cldc,kchd,ssmz,sras,zccs}. Another kind of protocols using bright pulses may have high successful entangling rate but  low initial entanglement fidelity\cite{llsy, llnm, wmrm}. Does there exist a scheme through which we may have both high successful entangling rate and high initial fidelity of entanglement?

Motivated by this consideration, we propose a protocol for efficient long-distance quantum
communication. In this scheme the entanglement between two qubits
assisted by high-Q  microtoroidal resonators in the basic segments
can be generated with a success probability near 0.5 by adopting
interference effect and state projection from measurements
\cite{ccgz}. The entanglement is then extended to a double length
through entanglement swapping. In both of the two stages, the
detection of qubit states is accomplished with high accuracy by
measuring many photons using single-photon detectors. From this
scheme the efficient built-in entanglement purification and
robustness against experimental imperfections can be achieved. The
obtained entanglement can be directly applied to perform
entanglement-based quantum communication protocols like quantum
teleportation, cryptography, and loophole-free Bell inequality test.
The communication time increases near linearly rather than
polynomially with the distance. For a distance of 2500 km it may
significantly decrease to about only 30 ms, by 7 orders of magnitude
smaller than that through DLCZ protocol. This is in the same order of magnitude
as the time scale for the light traveling over this distance. This scheme can produce results comparable to those in \cite{llsy} but avoid its complex entanglement purification process.  Recent
advances in the observation of strong interaction between photons
and single atoms through microscopic optical resonators
\cite{adwb,dpao} lay the foundation for this scheme.

\begin{figure}
\includegraphics[scale=0.3]{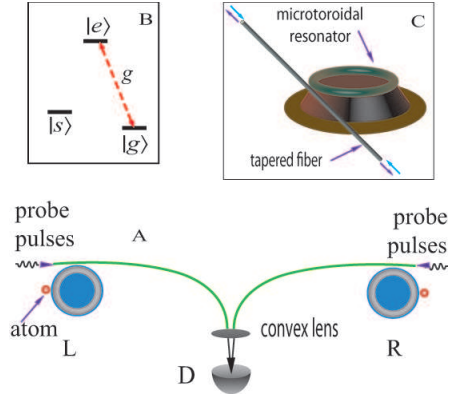}
\caption{\label{fig:1} (A) Schematic illustration of entanglement
establishment between two atoms L and R assisted by microtoroidal
resonators. (B) The level diagram of three-level $\Lambda$ atoms.  (C) schematic illustration of a microtoroidal resonator
with two modes.}
\end{figure}

The schematic description of the generation of  entanglement between
two qubits in a basic segment is shown in Fig. \ref{fig:1}. A
microtoroidal resonator has two internal counterpropagating modes
$a$ and $b$ with a common frequency $\omega_c$. These two modes are
coupled owing to the scattering \cite{adwb,dpao}. The intracavity
fields are coupled to a tapered fiber with high efficiency
$\varepsilon>0.99$ \cite{skpv}. The evanescent intracavity fields
 coherently interact with the ground state $|g\rangle$ and the
excited state $|e\rangle$ with energy $\omega_e$ of a three-level
atom which is near the external surface of the resonator with
well-defined azimuthal phase $\theta=\pi/2$ \cite{dpao}. A
metastable $|s\rangle$ of the atom does not interact with the field
of modes $a$ and $b$ due to being off resonance with the field. Such
a $\Lambda$ system may be provided by a donor atom in doped silicon
basis \cite{beka}, where the qubit states $|g\rangle$ and
$|s\rangle$ are encoded onto electron Zeeman sublevels and the
excited state is provided by the lowest bound-exciton state, or
other examples such as the hyperfine structure of trapped ions
\cite{llnm}.

In this paper we consider the situation where the input field and
the resonator are impedance-matched (critical coupling), which can
be reached through careful choice of the point of contact between
the surface of the microtoroid and the tapered optical fiber and
applying the input probe pulse with frequency $\omega_p=\omega_c$
\cite{adwb,dpao}. Under this condition, there are two cases in the
forward output of the tapered fiber: First, the output will drop to
zero due to the interference between the cavity field $a$ and the
input field $a_{in}$ when the atom is in state $|s\rangle$ (dark
state); Second, many single-photons come through the resonator one
by one with average interval time $\tau_B$ \cite {dpao}, when the
atom is in state $|g\rangle$ (bright state) and is in resonance
with the cavity $\omega_e=\omega_c$.

The qubits in nodes L and R are initialized in state
$(|g\rangle+e^{i\phi_i}|s\rangle)/\sqrt{2}$, $(i=L,R)$, then two same
weak probe pulses with frequency $\omega_p=\omega_c$ are incident
simultaneously into the resonators (Fig. \ref{fig:1}). If both of
the pulses contain only a single photon, then the state of the
system comprising of the atom and the forward output photon can be
described by
$|\psi\rangle_i=(|s\,0\rangle_i+a_i^\dagger|g\,0\rangle_i)/\sqrt{2}$,
$(i=L,R)$. The forward propagating photons are combined through such as a convex lens and measured at the
midpoint of the optical length between nodes R and L. The mode of
output photon measured by the single-photon detector D is
$a_e=(a_L+e^{i\varphi}a_R)/\sqrt{2}$, where $\varphi$ is an unknown
difference of the phase shift in the left and the right side
channels and can be set to zero \cite{cldc}. A click in detector D
measures the photon $a_e^\dagger a_e$ \cite{dlcz}. By applying $a_e$
to the state $|\psi\rangle_L\otimes|\psi\rangle_R$, we have the
projected state of the atoms L and R: $ |\psi(\phi)\rangle_{LR}=(|gs\rangle_{LR}+e^{i\phi}|sg\rangle_{LR})/\sqrt{2}                                                                                       $
with $\phi=\phi_L-\phi_R$ and the conventional notation
$|gs\rangle_{LR}= |g\rangle_L\otimes|s\rangle_R$. Considering the
presence of noise from the dark count of the photon detectors
described by a ``vacuum" coefficient $c_0$, the projected state
becomes $ \rho(c_0,\phi)=\frac{1}{c_0+1}(c_0|00\rangle_{LR}\langle00|+
    |\psi\rangle_{LR}\langle\psi|)$ \cite{dlcz}.

\begin{figure}
\includegraphics[scale=0.4]{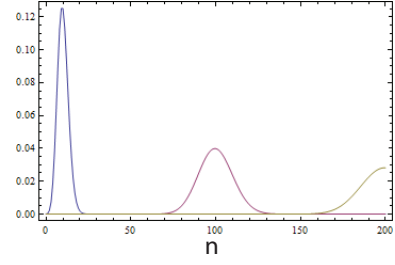}
\caption{\label{fig:2} The photons' Poisson distribution with
average photon number $\langle n\rangle=10$, $100$, and $200$,
respectively}
\end{figure}

If the probe pulses of duration $t_e$ contain many photons, we will
detect about $N=e^{-\frac{L_0}{2L_{att}}}t_e/\tau_B$ single-photons
arising from one atom bright, where $L_0$ is the length of a basic
link, $L_{att}$ is the channel attenuation length. These extra
single-photons except the first one will further confirm the state
of the system of atoms R and L, thus  reduce the error count
probability. For example, if $N_1=100$, according to \cite{dpao},
the number of background $N_b$ photons is only about $0.07N_1$, we
may assume $N_b=10$, and the number of photons from two atoms bright
will be about $N_2=200$. Assuming the probability of finding $n$
photons from $i$ $(i=0,1,2)$ atoms bright is given by a Poisson
distribution \cite{dpao,sblb} (see Fig. \ref{fig:2}), we find that
the probability $P_0$, $P_1$, and $ P_2$ of finding $n \in [40,120]$
photons due to zero, one, and two atoms bright is
$7.3\times10^{-13}$, 0.9773, and $6.7\times10^{-10}$, respectively.
Thus, we can unambiguously distinguish the state of one atom bright
from the other two states. The dark count in this case is excluded
and the projected state is $|\psi(\phi)\rangle$ provided that the
recorded photon number $n\in[40,120]$. As for the lost part of the
outgoing photons, since they are immediately lost, we can assume
that the subspace U where the lost photons enter is approximately
vacuum at all times (Born approximation) \cite{llnm}. Thus the whole
state can be written as
$\rho'=|\psi\rangle\langle\psi|\otimes|0\rangle\langle0|_U$. After
tracing the lost photons, we find that the state of the system of
atoms R and L can be described by $|\psi(\phi)\rangle$.

\begin{figure}
\includegraphics[scale=0.3]{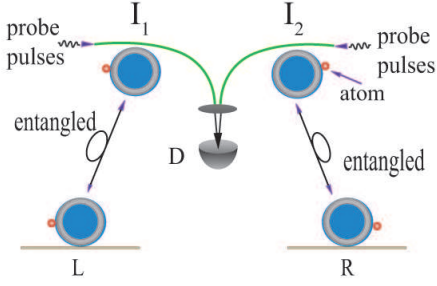}
\caption{\label{fig:3} Configuration for entanglement swapping.}
\end{figure}

After establishing entanglement within the basic links, the entanglement can
be extend to double length by entanglement swapping with the setup
schematically illustrated in Fig. \ref{fig:3}. Two pairs of
atom-resonator systems, L-I$_1$, and I$_2$-R, are prepared in states
$|\psi(\phi_1)\rangle$ and $|\psi(\phi_2)\rangle$, respectively. Two identical
probe pulses of duration $t_s=100 \tau_B$ simultaneously come into
the microtoroidal resonators $I_1$ and $I_2$, respectively. The
forward propagating photons are combined together and then detected
by the single-photon detector D at the midpoint of the optical
length between atoms I$_1$ and I$_2$. If and only if D records $n\in
[40,120]$ photons, i.e., one of the atoms I$_1$ and I$_2$ is bright,
the entanglement swap succeeds with the success probability
$p_1=0.5P_1$. Otherwise, the entanglement fails to extend, and the
previous entanglement generating and swapping process needs to be
repeated, till the protocol is successful at last. The detector D
measures $a_s^\dagger a_s$ with $a_s=(a_{I_1}+a_{I_2})/\sqrt{2}$.
Successful measurement will project the state
$|\psi(\phi_1)\rangle|\psi(\phi_2)\rangle$ into
$|\Psi(\phi)\rangle=(|gsgs\rangle_{LI_1I_2R}+e^{i\phi}|sgsg\rangle_{LI_1I_2R})/\sqrt{2}$
with $\phi=\phi_1+\phi_2$. Then, atoms $I_{1,2}$ are manipulated to
make a unitary transform $|s\rangle_{I_{1,2}}\rightarrow
|e\rangle_{I_{1,2}}$ by applying $\pi$ pulses and decay back to the
ground state $|g\rangle_{I_{1,2}}$ due to spontaneous emission.
Finally, we obtain the state $|\psi(\phi)\rangle_{LR}$. This
protocol for entanglement swap can be repeated to extend the
communication length. We have the success probability $p_i=0.5P_1$,
$(i=1,2,...,m)$ for the $i$th entanglement swap. Considering the
time for the signal traveling from D to L and R (see Fig.
\ref{fig:1}) to tell the controller whether or not to start the next
process, the average total time required for successful distribution
of entanglement over distance $L_t=L_m=2^mL_0$ is
  \begin{equation}\label{eq6}
T=\frac{T_0}{\prod_{i=0}^mp_i}=\frac{2^{m+1}T_0}{P_1^{m+1}},
\end{equation}
where $T_0= L_0/c +t_ee^{L_0/L_{att}}$ is the time needed to establish entanglement within the basic links with $c$ being the light
speed in the optical fiber. Note that $T$ increases near linearly
with the channel length.

\begin{figure}
\includegraphics[scale=0.3]{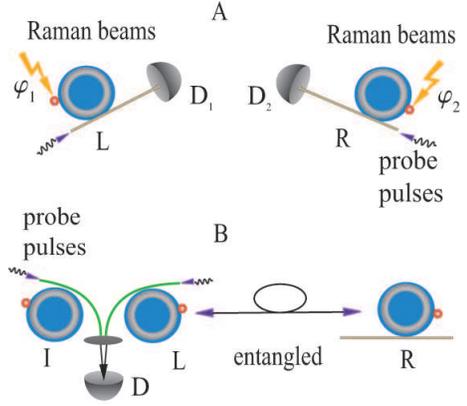}
\caption{\label{fig:4} (A) Configuration for the realization of
quantum cryptography and Bell inequality test. (B) Configuration for
quantum teleportation.}
\end{figure}

With the entangled state $|\psi(\phi)\rangle$ between two distant
sites in hand, we can apply the entanglement to quantum
communication protocols, such as quantum teleportation,
cryptography, and Bell inequality test directly. For  quantum
cryptograph and Bell inequality test (see Fig. \ref{fig:4}A), we
first make a local phase shift to transform the state
$\psi(\phi)\rangle_{LR}$ to state  $
    |\psi\rangle_{LR}^-=|gs\rangle_{LR}-|sg\rangle_{LR}$ \cite{dlcz}.
Then, two Raman beams are simultaneously applied to atoms L and R to
make $\phi_L$ and $\phi_R$ rotation about axis X, respectively.
Finally, two probe pulses of duration $t_a=100\tau_B$ are applied
simultaneously to the resonators to measure the states of atoms R
and L with the results 0 for the forward propagating photon number
$n<40$ recorded by detector $D_{1,2}$ and 1 for $n\geq40$. According
to the Ekert protocol for quantum cryptography \cite{aeke}, $\phi_L$
and $\phi_R$ are chosen randomly and independently from the set
$\{0, \pi/2\}$, the measurement results becomes the shared secret
key if the two sides get information through the classic
communication that they have chosen the same rotation. For the Bell
inequality test, we obtain the correlations
$E(\phi_L,\phi_R)=
\frac{N_s(\phi_L,\phi_R)-N_d(\phi_L,\phi_R)}{N_s+N_d}
=\cos(\phi_L-\phi_R)$ \cite{rkms},
where $N_s$ ($N_d$) denotes the number of measurements with two same
(different) results, and the result
$ E(0,\frac{\pi}{2},\frac{\pi}{4},\frac{3\pi}{4})=|E(0,\frac{\pi}{4}) +
E(\frac{\pi}{2},\frac{\pi}{4})+E(\frac{\pi}{2},\frac{3\pi}{4})
-E(0,\frac{3\pi}{4})|=2\sqrt{2}$
violates the CHSH inequality $E(0,\frac{\pi}{2},\frac{\pi}{4},
\frac{3\pi}{4})\leq2$ \cite{chsh}. Note that the outcome of every
Bell inequality experiment is used due to the unity state detection
efficiency. Combined with the space-like measurements, this scheme
can be used for a loophole-free test of the Bell inequality
\cite{rkms}.

To faithful transfer unknown quantum states $\varphi_I=\alpha
|g\rangle_I+\beta |s\rangle_I$ with arbitrary complexes $\alpha$ and
$\beta$ satisfying $|\alpha|^2+|\beta|^2=1$, we can use quantum
teleportation protocol through the entangled state
$\psi(\phi)\rangle_{LE}$ (see Fig. \ref{fig:4}B). Two identical
probe pulses of duration $t_t=100\tau_B$ simultaneously enter the
microtoroids, the forward propagating photons are mixed together and
measured by the single-photon detector D, which records the photon
$a^\dagger_t a_t$ with $a_t=(a_I+e^{i\phi}a_L)/\sqrt{2}$. If the detector
records $n\in[40,120]$ photons, the protocol succeeds with a
probability of $0.5P_1$ and the state
$|\psi\rangle_I|\varphi\rangle_{LR}$ is projected into a state
$\alpha|gsg\rangle_{ILR}+\beta|sgs\rangle_{ILR}$. Then, atom I and L
are manipulated  to make a unitary transformation
$|s\rangle_{I,L}\rightarrow |e\rangle_{I,L}$ and decay to the ground
state $|g\rangle_{I,L}$. Finally, we obtain the state
$|\varphi\rangle_R=\alpha |g\rangle_R+\beta |s\rangle_R$.

Because of the unambiguous detection of the state of qubits with unity detection efficiency  through
measuring many single-photons rather than one arising from the atom
bright, the errors from, e.g., the dark count, low detection efficiency, and the imperfection
that the single-photon detector may not distinguish between one and
two photons can be overcome in this scheme. This results in the
increase of the quantum communication speed and the corresponding
fidelity. Under the Born approximation, the influence arising from
the lost photons in the optical fiber and from the spontaneous
emission in stages of entanglement extension and teleportation may
be negligible. Owing to the limited coherence time $t_c$, atoms may
decay from metastable state $|s\rangle$ to the ground state
$|g\rangle$. To solve this problem, we may store the spin state in
nuclear memory which has the longest decoherence time in all quantum
systems so far. Fast electron-nuclear double resonance (ENDOR) pulse
techniques may be used  for prompt storage and retrieve of the
electron-spin state \cite{jgpd,llsy}. If two atom $I_1$ and $I_2$ in Fig.\ref{fig:3} decay from $|s\rangle$ to $|g\rangle$, detection D measures $n>120$ photons, thus, the entanglement swap fails. In a similar way, we can show that in the whole process of distributing entanglement over long distance, there
may exist only one atom decaying with at most probability $0.5P_1(1-
e^{-t_s/t_c})$ that we cannot exclude through detection of $n$
photons to ensure  only one atom bright, except for the two atoms
apart with a distance $L_t$. The decoherence arising from the
spontaneous decay  can be written as $\Delta F<3(1-e^{-t_e/t_c})$.
Assuming $L_t=2^6 L_0 =2500$ km, $c=2.0\times10^{8}$ m/s,
$L_{att}=22$ km \cite{ssmz}, from equation \ref{eq6}, we have $T=30$
ms, which is on the same order of the time $L_t/c=12.5$ ms. Compared
with the corresponding time $T=650000$ s using the DLCZ protocol for
expected fidelity $F=0.9$ provided that many atoms excitations are
the only imperfection in the experiment \cite{ssmz}, the total
average time for successful distributing entanglement over that
distance can be reduced by 7 orders of magnitude. Assuming
$\tau_b=6$ ns \cite{dpao}, $t_c=6$ ms, we have $\Delta F<0.0018$.

Stationary and non-stationary phase shifts from stationary and
non-stationary channels and set-up asymmetries, respectively, are
main decoherence source in our scheme. The non-stationary phase
shifts increase with the length by the random-walk rule $\sqrt{L_m/L_0}$  and can be
reduced to a negligible degree \cite{dlcz}.
Because the total average time $T$ decreases significantly by
several orders of magnitude, this phase shifts in our scheme  can be
overcome much easier than that in DLCZ protocol \cite{zccs, hhjy}.
The stationary phase shifts are easier to handle than the
non-stationary ones are. The mechanism for photon blockade used in
our scheme is very robust against many experimental imperfections
\cite{dpao}.

As a summary, we present a microtoroidal resonator-based scheme for
efficient long-distance quantum communication. Moreover our scheme shows high fidelity and
good robustness against many experimental imperfections. To trap single atoms near the surface of the microtoroidal
resonator is still a technical challenge. However, with the rapid
advances in the relevant technologies it is no doubt that in near future the obstacle will be overcome. This scheme may open
up realistic probability of efficient long-distance quantum
communication.

\section {ACKNOWLEDGMENTS}
This work was supported by the State Key Programs for Basic Research
of China (2005CB623605 and 2006CB921803), and by National Foundation
of Natural Science in China Grant Nos. 10474033 and 60676056.

\end{document}